\documentclass[twocolumn,aps]{revtex4-1}
\usepackage{graphicx}

\begin{document}
\title{Relationship between the transverse-field Ising model and the XY model via the rotating-wave approximation}
\author{Thomas G. Kiely}
\author{J. K. Freericks}
\affiliation{Department of Physics, Georgetown University, 37th and O Streets NW, Washington, DC 20057, USA}
\begin{abstract}
In a large transverse field, there is an energy cost associated with flipping spins along the axis of the field. This penalty can be employed to relate the transverse-field Ising model in a large field to the XY model in no field (when measurements are performed at the right stroboscopic times). We describe the details for how this relationship works and, in particular, we also show under what circumstances it fails. We examine wavefunction overlap between the two models and observables, such as spin-spin Green's functions. In general, the mapping is quite robust at short times, but will ultimately fail if the run time becomes too long. There is also a trade-off between the length of time one can run a simulation out to and the time jitter of the stroboscopic measurements that must be balanced when planning to employ this mapping. 
\end{abstract}
\maketitle

\section{Introduction}

Recently, there has been significant activity in employing the transverse-field Ising model within quantum simulators to examine adiabatic state preparation, excitation spectroscopy, quantum propagation speeds (Lieb-Robinson bounds) and complicated many-body phenomena like many body localization and time crystals~\cite{monroe_3,monroe_10,bollinger_j,monroe_af,monroe_spectra,monroe_lr,rainer_lr,bollinger_otoc,
monroe_time,monroe_mblocal,monroe_53}. The transverse-field Ising model is given by
\begin{equation}
\mathcal{H}_{TFI}=\sum_{i<j}J_{ij}\sigma_i^x\sigma_j^x-B\sum_i\sigma_i^z
\label{eq: tfi_ham}
\end{equation}
where $\sigma_i^\alpha$ is the Pauli spin matrix at site $i$ in spatial direction $\alpha$. The spin-spin interactions are given by $J_{ij}$ for the interaction between spins at sites $i$ and $j$ and will be called the spin-exchange piece of the Hamiltonian $\mathcal{H}_{TFI}^{SE}$, while the magnetic field strength in the $z$-direction is given by $B$ (and the magnetic field piece of the Hamiltonian is denoted $\mathcal{H}_{TFI}^B$). In an adiabatic state preparation, the system would be initialized in a state polarized along the $z$-direction and then the field would be slowly reduced in the presence of the $J_{ij}$ until the system evolved into the ground state of the Ising model with no field. If the system is evolved too rapidly, then diabatic excitations will occur, and their energies can be measured via different spectroscopy techniques. Lieb-Robinson bounds~\cite{lieb_robinson} can be inferred by measuring the propagation speeds of disturbances to the spin chain, while many body localization and time crystals require somewhat more sophisticated arrangements that include quasi disorder added to the system.

There also is an interest in going beyond the simple transverse-field Ising model to more complex systems. Here, one can imagine going to more complex spin models, like the XY model or the Heisenberg model, or one can imagine going to higher spin representations, like going to spin one instead of spin one-half. In this paper, we will focus on employing the rotating-wave approximation to go from the transverse-field Ising model to the XY model, which is given by the following Hamiltonian:
\begin{equation}
\mathcal{H}_{XY}=\frac{1}{2}\sum_{i<j}J_{ij}(\sigma_i^x\sigma_j^x+\sigma_i^y\sigma_j^y)
\label{eq: xy_ham}
\end{equation}
While it may not seem obvious, there is a rotating-wave approximation approach which will allow us to map the transverse-field Ising model into the XY model. We describe this next.

Define the spin raising and lowering operators via $\sigma^{\pm} =\sigma^{x} \pm i \sigma^{y}$. Inverting these relations lets us write $\sigma^x=(\sigma^++\sigma^-)/2$ and $\sigma^y=(\sigma^+-\sigma^-)/2i$. A quick calculation then shows that $\sigma_i^x\sigma_j^x+\sigma_i^y\sigma_j^y=(\sigma_i^+\sigma_j^-+\sigma_i^-\sigma_j^+)/2$. Thus, the XY model can be represented in terms of these raising and lowering operators. To find the relationship between the transverse-field Ising model and the XY model, we substitute the raising and lowering operators into the transverse-field Ising model in Eq.~(\ref{eq: tfi_ham}), by replacing $\sigma^x$ by ($\sigma^++\sigma^-)/2$ everywhere. This yields
\begin{equation}
\mathcal{H}_{TFI}=\frac{1}{4}\sum_{i<j}(\sigma_i^+\sigma_j^++\underbrace{\sigma_i^+\sigma_j^-+\sigma_i^-\sigma_j^+}_{\rm XY~piece}+\sigma_i^-\sigma_j^-)-B\sum_i\sigma_i^z.
\label{eq: tfi_ham2}
\end{equation}
Note how two of the spin-exchange terms are the same as those in the XY model, but there are two other operators which involve either raising the spins twice or lowering them twice.  If the field $B$ is large, there will be a large energy cost for those double spin flips, as opposed to having no energy cost for the XY terms which flip one spin up and the other spin down. This provides a hint that there should be a relationship between these two models in a large magnetic field.

To make the mapping more precise, we will invoke the rotating-wave approximation, which requires us to go to the interaction representation with respect to the magnetic field piece of the Hamiltonian (or, equivalently, to the rotating frame). We then transform the spin-exchange part of the Hamiltonian via $\mathcal{H}_{TFI}^{SE}\rightarrow \exp[i\mathcal{H}_{TFI}^Bt]\mathcal{H}_{TFI}^{SE}\exp[-i\mathcal{H}_{TFI}^Bt]$ to give us the interaction representation of the ``perturbation.''  In this rotating frame, the magnetic-field piece of the Hamiltonian is accounted for in the time dependence under the ``unperturbed Hamiltonian'' $\mathcal{H}_0$, so the ``interaction piece'' of the transverse-field Ising model becomes
\begin{equation}
\mathcal{H}_{TFI}\rightarrow
\frac{1}{4}\sum_{i<j}(\sigma_i^+\sigma_j^+e^{4iBt}+\sigma_i^+\sigma_j^-+\sigma_i^-\sigma_j^++\sigma_i^-\sigma_j^-e^{-4iBt}).
\label{eq: tfi_rot}
\end{equation}
When the magnetic field, $B$, is large, the rotating terms create rapidly oscillating terms in the Hamiltonian which average to zero and can be ignored via the rotating-wave approximation. So, whenever the rotating-wave approximation can be applied to the transverse-field Ising model, it should act like an XY model.
To fully understand this mapping, though, we need to explore in detail how the energy eigenvalues relate as well as the dynamics of the wavefunctions. In doing so, we will find the mapping holds stroboscopically in time because the frame is rotating at the Larmor frequency, and hence the wavefunctions have an oscillating phase which returns to a multiple of $2\pi$ every Larmor period. We will need to balance the improved accuracy given from a larger field with the difficulty in properly timing the stroboscopic measurements when the complex phase factors oscillate too rapidly.

In Sec. II, we derive the formalism we employ for making these comparisons. In Sec. III, we present results that illustrate both the success of the mapping and also show under what circumstances it fails. This is followed up by conclusions in Sec. IV. 

\section{Formalism}
\subsection{Spin Exchange Coefficients}

In an ion trap, the internal ion states of a given atomic species are mapped onto the spins of a two-state system. A spin-dependent force is applied to the system, and in the situation where the phonons are only virtually created, they can be adiabatically eliminated from the system producing an effective spin-spin interaction. The spin-spin couplings vary with time, but their average values are given by~\cite{monroe_spin_exchange}
\begin{equation}
J_{ij} = \Omega^{2} \omega_{R} \sum_{m=1}^{N} \frac{b_{i,m}b_{j,m}}{\mu^{2} - \omega_{m}^{2}},
\end{equation}
where $\Omega$ is the Rabi frequency, $\omega_{R}$ is the atomic recoil frequency, $b_{i,m}$ are the transverse phonon normal modes of the ion chain (labeled by the mode index $m$ and the spatial position $i$), $\omega_{m}$ are the corresponding normal mode frequencies, and $\mu$ is the detuning frequency. The normal modes $b_{i,m}$ and normal-mode frequencies $\omega_{m}$ are found from a straightforward classical mechanics calculation once the trap parameters are known~\cite{james1,james2}. The highest frequency transverse normal mode is the center-of-mass (COM) mode. When the detuning is larger than the COM mode frequency $\mu > \omega_{COM}$, the spin-exchange coefficients $J_{ij}$ are well-approximated by
a simple power law
\begin{equation}
J_{ij} \approx \frac{J_{0}}{|i-j|^{\alpha}},
\end{equation}
where $\alpha$ varies from $0$ to $3$ depending on the parameters of the Paul trap and the detuning.
All frequencies in this paper that are expressed in units of Hz are regular frequencies; the corresponding angular frequencies are $2\pi$ times larger.  We use the trapping parameters of a recent experiments~\cite{monroe_lr}: $\Omega\sqrt{\omega_{R}/\omega_{trans}} = 20$~kHz, $\omega_{trans} = 4.80$~MHz, and $\mu = \omega_{COM}+ 60$~kHz, where $\omega_{COM}$ is the transverse center of mass phonon mode of the ion chain and is equal to $\omega_{trans}$. We controlled the exchange coefficients by varying the anisotropy of the trap, that is, the ratio of the longitudinal to the transverse trapping frequency.  We keep $\omega_{trans}$ fixed and vary $\omega_{lon}$ from $ 560-950$~kHz, which yields an $\alpha$ varying between $0.63$ and $1.19$ with $J_{0} \approx  500$~Hz.

\subsection{Time Evolution}
Both the transverse-field Ising model and the XY model are time-independent. The evolution operator is then given by
$U(t)=\exp(-i\mathcal{H}t)$.
If $U(t)$ is acting on a state that is not an energy eigenstate, then it is convenient to diagonalize the Hamiltonian in the exponential using $V$, a unitary matrix whose rows are the eigenvectors of $\mathcal{H}$, so that
\begin{equation}
U(t)| \Psi \rangle = V^{\dagger} e^{-i V \mathcal{H} V^{\dagger} t} V | \Psi \rangle.
\end{equation}
Since we work in the same basis for both the transverse-field Ising model and the XY model, their respective evolution operators acting on a single initial state provides a direct comparison between the evolved states.

\subsection{Energy Levels}
\begin{figure}
\includegraphics[width=3.2in]{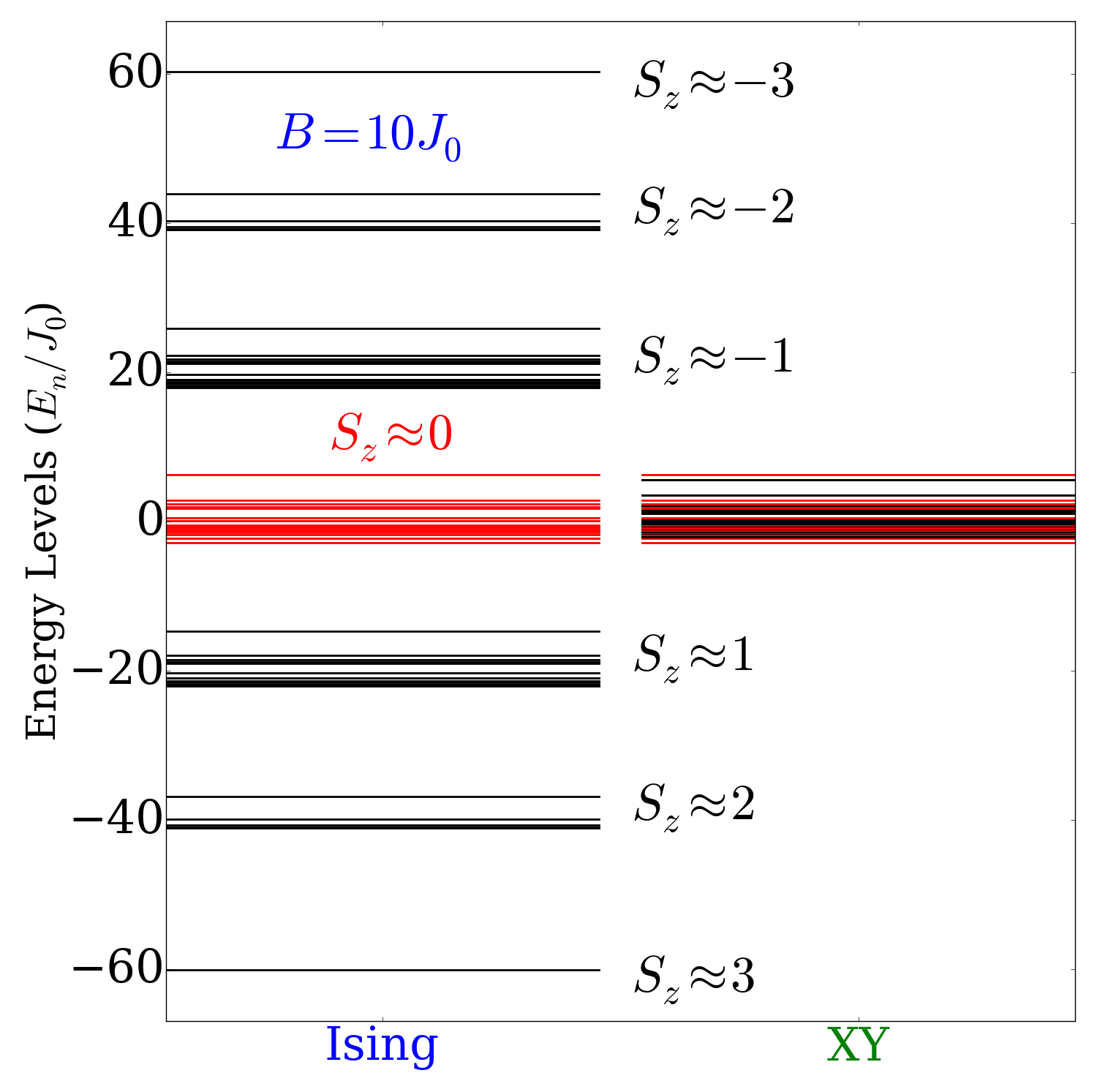}
\caption{(Color online.) Energy levels of the transverse-field Ising Hamiltonian in a field of $B/J_{0}=10$ and of the XY Hamiltonian for a chain of $6$ ions and a longitudinal trapping frequency of $ 950$~kHz and $\alpha\approx 0.63$. The organization of the transverse-field Ising levels into approximate $S^{z}_{tot}$ blocks is evident. The $S^{z}_{tot}\approx 0$ Ising levels and the $S^{z}_{tot}=0$ XY levels are colored in red.}
\label{fig:macroSpectra}
\end{figure}

Our first illustration of the mapping between these models involves a comparison of their energy levels. The transverse-field Ising model energy levels in a strong transverse field are approximately Zeeman shifted by $-2mB$, where $m$ is the eigenvalue of the $S^{z}_{tot}=\sum_i\sigma_i^z/2$ operator. The shift is approximate because $S^{z}_{tot}$ does not commute with the transverse-field Ising Hamiltonian. We identify approximate $S^{z}_{tot}$ blocks in the transverse-field Ising energy levels in the limit of a large transverse field; that is, the energy levels will split based on the approximate value of $S^{z}_{tot}$ acting on the corresponding eigenstate. Figure \ref{fig:macroSpectra} shows the extent to which this is possible when $B/J_{0}=10$ in a $6$-ion chain with $\omega_{lon}=950$~kHz and $\alpha\approx 0.63$. The XY Hamiltonian commutes with the $S^{z}_{tot}$ operator, so we can compare the energy states of both models on the basis of their $S^{z}_{tot}$ value (approximate for the transverse-field Ising model and exact for the XY model).

For systems with an even number of spin sites, we can directly compare the $S^{z}=0$ energy levels of the XY model to the $S^{z}\approx 0$ Ising energy levels, as those levels are not Zeeman shifted to linear order in $B$.

\begin{figure}
\includegraphics[width=3.5in]{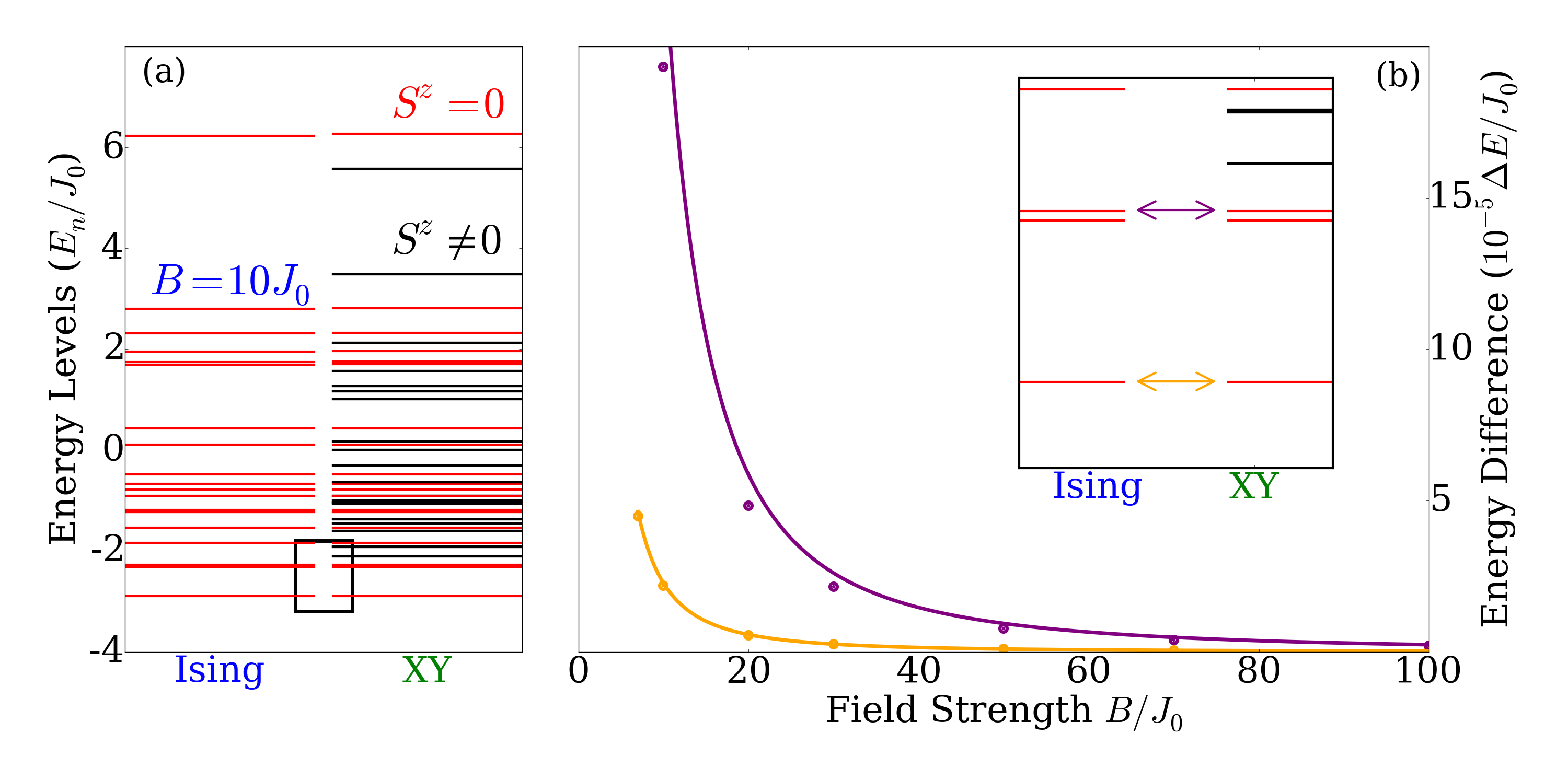}
\caption{(Color online.) (a) Transverse-field Ising model energy levels for $B/J_{0}=10$ and XY model energy levels for a chain of $6$ ions in a longitudinal trapping frequency of $ 950$~kHz ($\alpha\approx 0.63$). Levels in the $S^{z}_{tot} \approx 0$ block of the Ising model and the $S^{z}_{tot}=0$ block of the XY model are colored in red. (b) Difference between the corresponding XY and transverse-field Ising model energy levels is plotted as a function of field strength for $B/J_{0}=7$, $10$, $20$, $30$, $50$, $70$, and $100$. The particular levels used to measure the difference are shown in the inset with the arrows.}
\label{fig:spectra}
\end{figure}

The Ising energy levels in the limit of a large transverse field can be treated perturbatively, where the zero-field Ising Hamiltonian perturbs the transverse-field Hamiltonian. In a simultaneous eigenbasis of the $S^{2}_{tot}$ and $S^{z}_{tot}$ operators, the magnetic-field-only Hamiltonian is highly degenerate. Fortunately, the zero-field Ising perturbation completely lifts the degeneracy. We diagonalize the $S^{z}_{tot}$ blocks of the full transverse-field Ising Hamiltonian, and then sum over the contribution of other $S^{z}_{tot}$ blocks to calculate the perturbative corrections. The second-order perturbative correction is given by
\begin{equation}
E^{(2)}_{n}=\sum_{m \neq n} \frac{\langle m|\mathcal{H}_{TFI}^{SE}|n\rangle \langle n|\mathcal{H}_{TFI}^{SE}|m \rangle}{E^{0}_{n}-E^{0}_{m}},
\end{equation}
where  $E^{0}_{n}$ is the unperturbed energy of the eigenstate $|n\rangle$ of $\mathcal{H}_{TFI}^B$. Second-order corrections to the energies of the $S^{z}_{tot}\approx 0$ block are all equal to zero. This indicates that the energies of the transverse-field Ising Hamiltonian are even functions of $B$, as the energies in the denominator are linear in $B$. The third-order correction, is then given by
\begin{eqnarray}
E^{(3)}_{n}&=&\sum_{m \neq n} \sum_{m' \neq n} \frac{\langle n|\mathcal{H}_{TFI}^{SE}|m\rangle \langle m|\mathcal{H}_{TFI}^{SE}|m'\rangle \langle m'|\mathcal{H}_{TFI}^{SE}|n\rangle}{(E^{0}_{n}-E^{0}_{m'})(E^{0}_{n}-E^{0}_{m})}\nonumber\\
& -& \langle n|\mathcal{H}_{TFI}^{SE}|n\rangle \sum_{m \neq n} \frac{\langle m|\mathcal{H}_{TFI}^{SE}|n\rangle \langle n|\mathcal{H}_{TFI}^{SE}|m\rangle}{(E^{0}_{n}-E^{0}_{m})^{2}},
\end{eqnarray}
and is non-zero, which indicates that the $S^{z}_{tot}\approx 0$ Ising levels and $S^{z}_{tot}=0$ XY levels should approach each other as $1/B^{2}$. Figure \ref{fig:spectra} shows the calculated energy differences at various field strengths, as well as a fit from the third-order perturbative correction, for a $6$-ion chain with $\omega_{lon}=950$~kHz ($\alpha\approx 0.63$).

\begin{figure}
\includegraphics[width=3.5in]{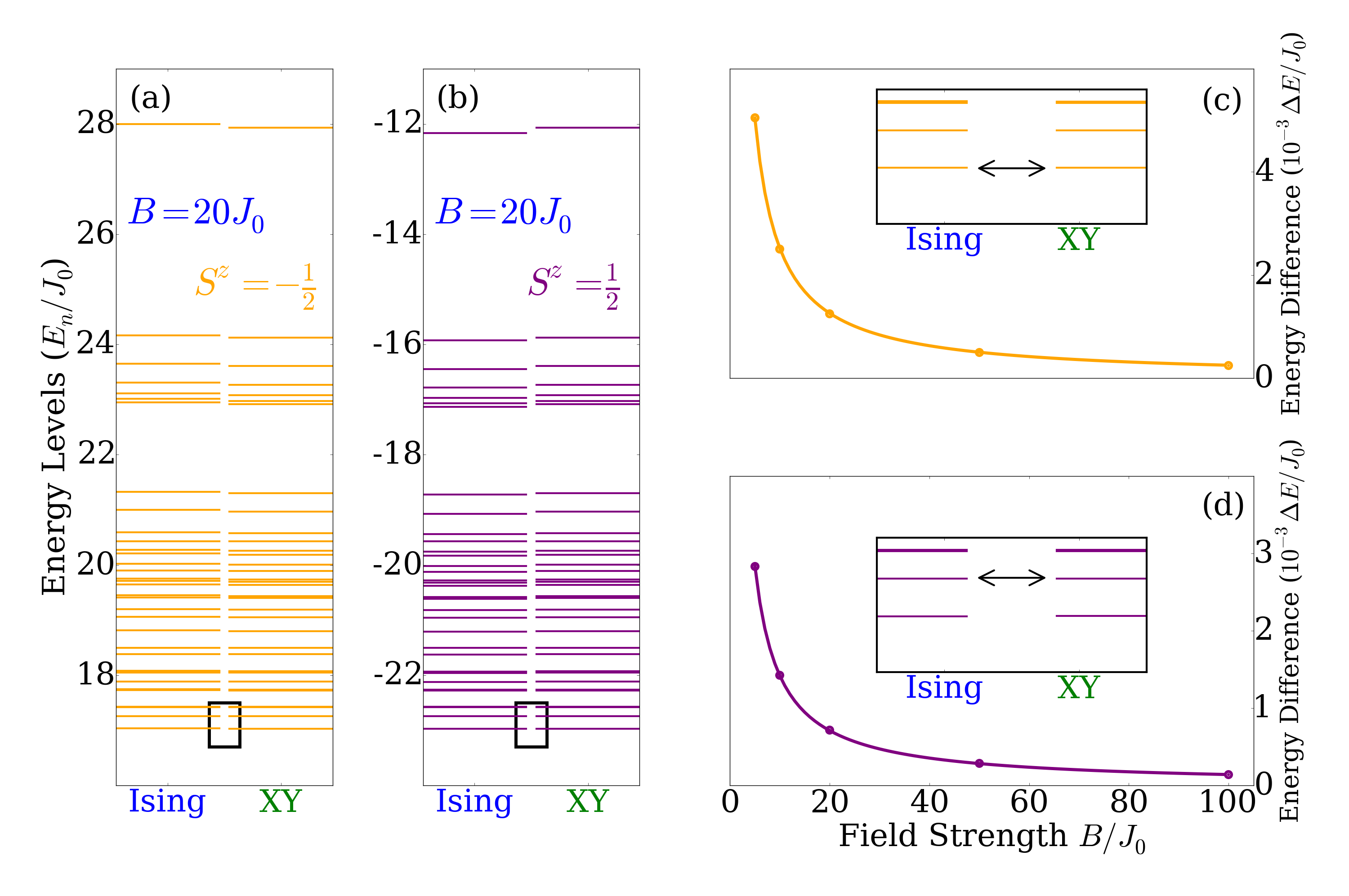}
\caption{(Color online.) (a) XY and transverse-field Ising model energy levels for $S^{z}_{tot}=-\frac{1}{2}$ in an external field of $B/J_{0}=10$ on a chain of $7$ ions with a longitudinal trapping frequency of $ 650$~kHz ($\alpha\approx 1$). The $S^{z}_{tot}=\frac{1}{2}$ energy levels are plotted in panel (b). Panels (c) and (d) plot the field dependence of the difference between $S^{z}_{tot}=-\frac{1}{2}$ and $S^{z}_{tot}=\frac{1}{2}$ energy levels, respectively. The levels used to calculate the differences are identified in the inset by the arrows.}
\label{fig:shiftedSpectra}
\end{figure}

For systems with an odd number of lattice sites, we need to shift the energy scales before comparing energy levels to account for the approximate Zeeman shift of a spin one-half state. Noting that $S^{z}_{tot}$ commutes with the XY Hamiltonian, adding a transverse magnetic field to the XY model will result in Zeeman shifts that are exactly linear in the field strength. We therefore compare the transverse-field Ising energy levels with $\mathcal{H}_{XY}^{B\neq 0} = \sum_{i<j} \frac{J_{ij}}{2} (\sigma^{x}_{i} \sigma^{x}_{j} + \sigma^{y}_{i} \sigma^{y}_{j}) -B \sum_{i} \sigma^{z}_{i}$ when both models have an equal field strength. This comparison is shown visually in Fig.~\ref{fig:shiftedSpectra} for a $7$-ion chain with $\omega_{lon}= 650$~kHz, where the fit in panels (c) and (d) goes as $1/B$. Second-order perturbative corrections to transverse-field Ising blocks with $S^{z}_{tot} \neq 0$ are nonzero, which explains why the correction for $S^{z}_{tot} \neq 0$ blocks no longer go as $1/B^{2}$.

\subsection{Wavefunction Overlap}
\begin{figure*}
\includegraphics[height=10cm]{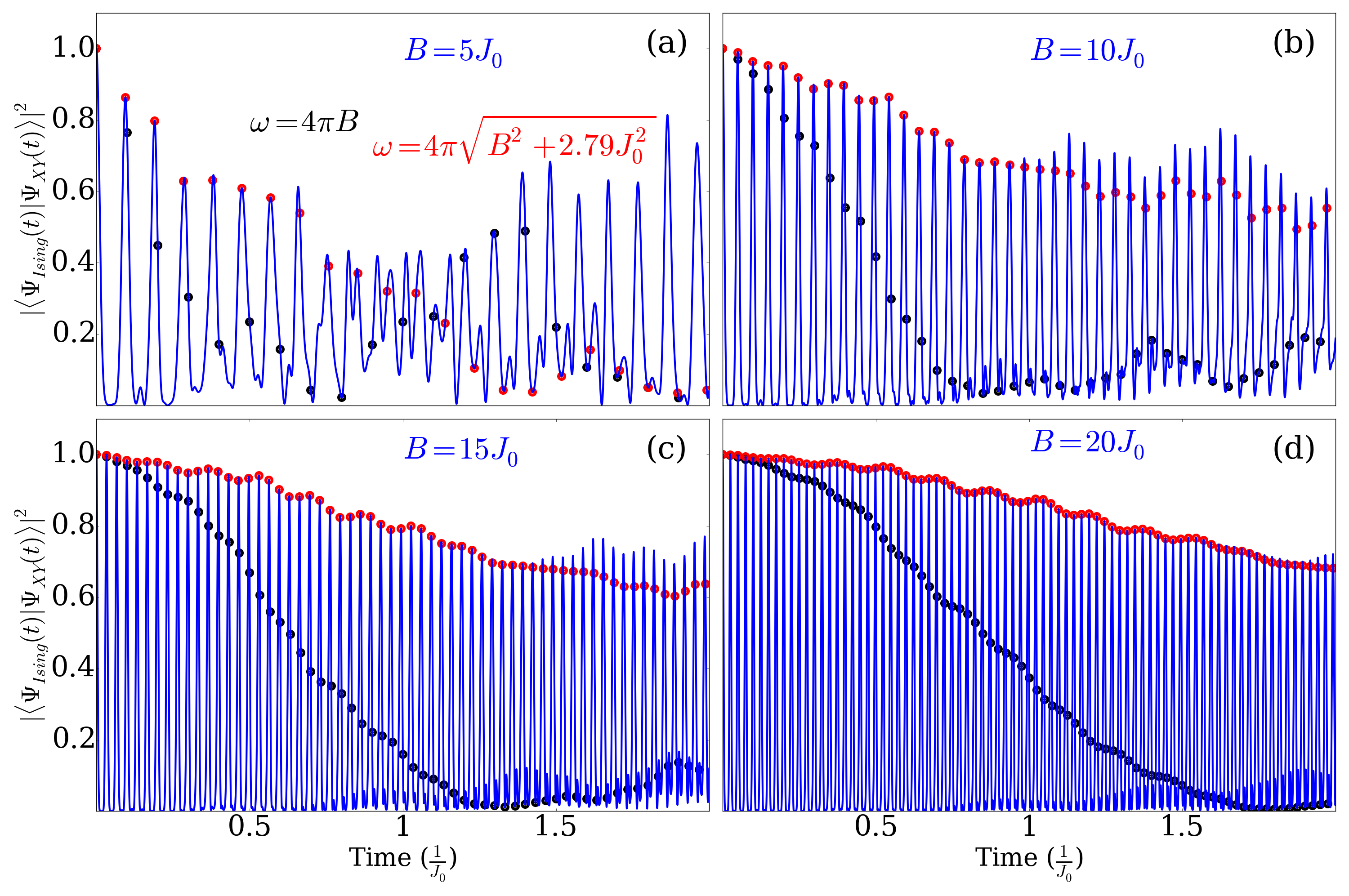}
\caption{(Color online.) Plot of the modulus squared of the overlap of the time-evolved XY and transverse-field Ising state vectors, $\langle \Psi_{Ising}(t)|\Psi_{XY}(t) \rangle$, for a $5$-ion chain with a longitudinal trapping frequency of $950$~kHz ($\alpha\approx 0.63$) and various transverse field strengths. Panels (a)-(d) plot the squared overlap between $tJ_{0}=0$ and $tJ_{0}=2$ for $B/J_{0}=5$, $10$, $15$, and $20$, respectively. The black dots are plotted at the Larmor frequency, $\omega_{L}=4\pi$. The red dots are plotted at a numerically-optimized frequency, given by $\omega_{opt} = 4\pi \sqrt{B^{2}+(1.67J_{0})^{2}}$.}
\label{fig:overlap}
\end{figure*}

A numerical evaluation of the modulus squared of the overlap between the XY and Ising wavefunctions as a function of time is shown in Fig.~\ref{fig:overlap} for a $5$-ion chain, $\omega_{lon}= 950$~kHz ($\alpha\approx 0.63$), and field strengths of $B/J_{0}=5$, $10$, $15$, and $20$. The initial state for these calculations is the state with all spins oriented in the $-\hat{y}$ direction, which is represented in the $z$-basis as the direct-product state $| \Psi \rangle = (|\uparrow\rangle - i|\downarrow\rangle)_{1} \bigotimes (|\uparrow\rangle - i|\downarrow\rangle)_{2} \bigotimes \ldots \bigotimes (|\uparrow\rangle - i|\downarrow\rangle)_{N}$. The black dots are placed at $2\pi n/\omega_L$, for integer $n$ with $\omega_L=4\pi B$, which corresponds multiples of the Larmor period. The red dots are placed according to an optimized frequency, which is found by modifying the Larmor frequency from $4\pi B$ to $4\pi \sqrt{B^{2}+(a J_{0})^{2}}$ and varying $a$ until the combined sum of all modulus squares of the overlaps for a given range of times reaches a local maximum at integer multiples of the modified period. The form of the correction assumes that the spin-exchange interaction can be treated as a mean field in the $\hat{x}$ direction and that the frequency of the oscillations depends on the resultant magnitude of the total field. The corrected frequency in Fig.~\ref{fig:overlap} corresponds to $a=1.67$, which was determined by optimizing the sum of all plotted points between $tJ_{0}=0$ and $tJ_{0}=1$. In general, we found that $a$ depends on the lattice size and on the initial state of the system, so it is not easy to know what it would be without solving the problem {\it a priori}. We went through this exercise to try to extend the period in time where the two models had wavefunctions that could be identified with each other stroboscopically. In general, however, if we don't have more accurate information available to us, we simply have to use the Larmor period, which breaks down a bit sooner than the corrected period.

The squared overlap oscillates between $1$ and $0$ at the Larmor frequency, while the envelope of the amplitude decays with time. The importance of the mean-field correction to the measurement frequency is shown clearly by the rapid rate of decay of the black dots relative to the red dots. Even for $B/J_{0}=20$, measurements taken with $\omega_{L}=4\pi B$ will fall so far out of phase by $tJ_{0}=1$ that $|\langle \Phi_{Ising}|\Phi_{XY} \rangle|^2\approx0.5$, even though the state vectors are still coming into a maximum alignment of $|\langle \Psi_{Ising}|\Psi_{XY} \rangle|^2\approx0.9$ at slightly different stroboscopic times.

\subsection{Green's Function}
We define the ``pure-wavefunction'' retarded spin-spin Green's function via
\begin{equation}
G^{R}_{\alpha, \beta, i, j}(t, t_{0}) = i \theta (t-t_{0}) \langle \Psi_{0} | [\sigma^{\alpha}_{i}(t), \sigma^{\beta}_{j}(t_{0})] | \Psi_{0} \rangle
\end{equation}
where $\sigma^{\alpha}_{i}(t) = U^\dagger(t) \sigma^{\alpha}_{i} U(t)$ is a Pauli matrix in the Heisenberg picture. The equilibrium Green's function (which would have a trace over all states rather than the pure-wavefunction definition above) can be easily shown to be invariant to translations in time, so that $G^{Req}_{\alpha, \beta, i, j}(t,t_{0}) = G^{Req}_{\alpha, \beta, i, j}(t+t',t_{0}+t')$.  In the wave function form, this is only the case when $| \Psi_{0} \rangle$ is an eigenstate of the Hamiltonian. Since we cannot choose an initial state which is an eigenstate of both the XY and transverse-field Ising Hamiltonians, this definition of the pure-wavefunction retarded Green's function is not always time-translation invariant. For transverse field strengths on the order of $10 J_{0}$ and times on the order of $\frac{1}{J_{0}}$, however, deviations of this Green's function from a time translation invariant one are negligible, so we ignore them. We choose to compare the $G^{R}_{x,x,i,j}$ components of the Green's function because they can be measured experimentally with Ramsey spectroscopy~\cite{demler1,demler2,yoshimura_freericks}.

\begin{figure*}
\includegraphics[height=10cm]{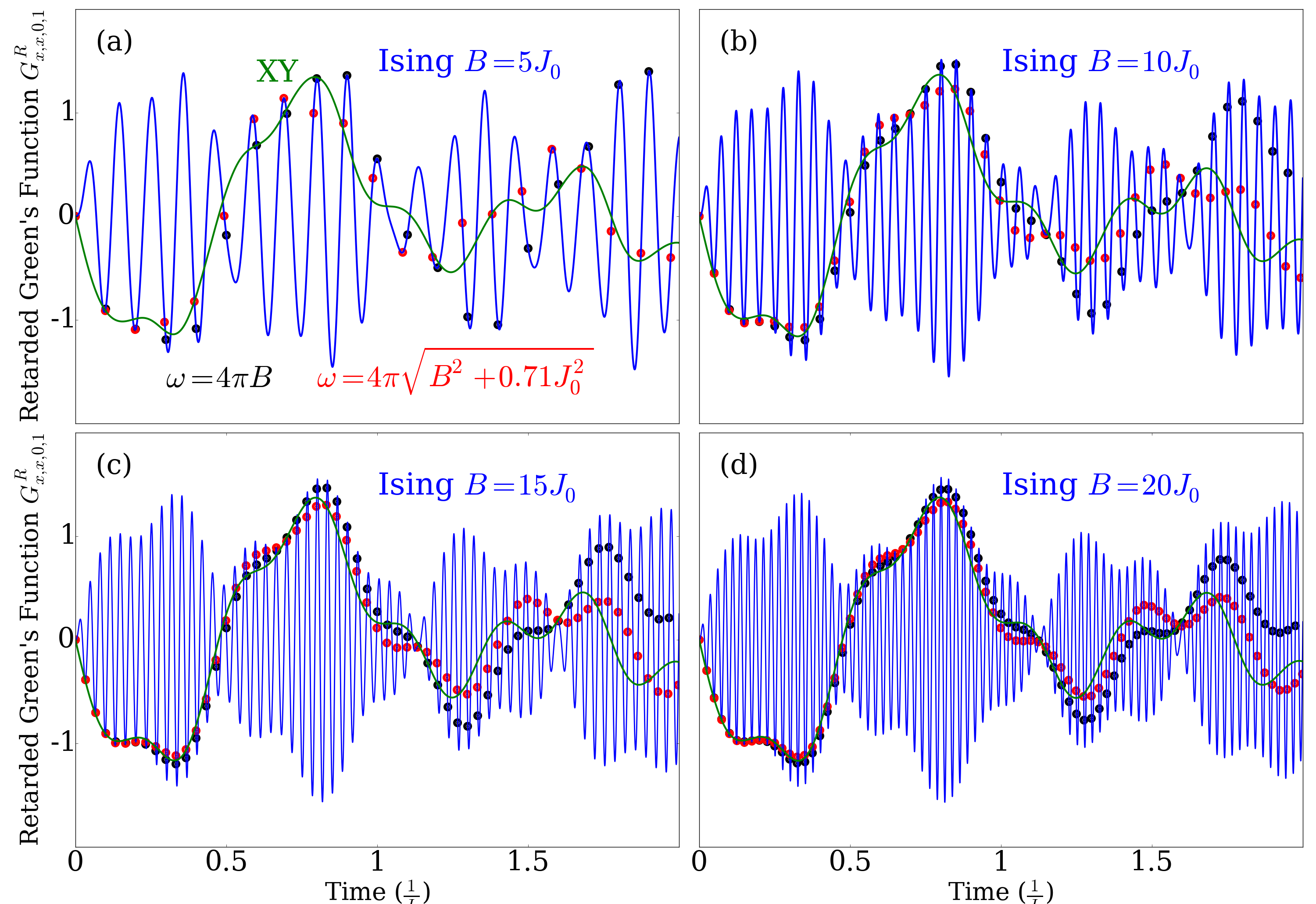}
\caption{(Color online.) Plot of the pure-wavefunction retarded Green's function of the transverse-field Ising Hamiltonian, in blue, and of the XY Hamiltonian, in green, for a chain of $7$ ions in a longitudinal trapping frequency of $650$~kHz ($\alpha\approx 1$) and various transverse field strengths. Panels (a)-(d) plot the Green's functions between $tJ_{0}=0$ and $tJ_{0}=2$ for $B/J_{0}=5$, $10$, $15$, and $20$, respectively. The black dots are plotted at the Larmor frequency, $\omega_{L}=4\pi$ and the red dots are plotted at a $\omega_{opt} = 4\pi \sqrt{B^{2}+(0.84J_{0})^{2}}$.}
\label{fig:greens}
\end{figure*}

In Fig.~\ref{fig:greens}, we show the numerical evaluation of the $G^{R}_{x,x,0,1}(t,0)$ for a $7$-ion chain in the XY model and the transverse-field Ising model with $B/J_{0}=5$, $10$, $15$, and $20$ and $\omega_{lon}= 650$~kHz ($\alpha\approx 1$). The pure state used in the Green's function calculation is defined by $|\Phi_{T}\rangle = \sum_{n} \sqrt{\frac{\exp[-\beta E_{n}]}{Z}} |n\rangle$, where $Z=\sum_{n} \exp[{-\beta E_{n}}]$ and $\beta=\frac{2}{J_{0}}$. Note that this wavefunction is not a thermal state, but it is a linear combination of the eigenstates with the amplitudes of each state chosen to have the same probability as in a thermal state~\cite{lim}. Dots indicate measurements of the transverse-field Ising model Green's function at the particular times which correspond to the simulation of the XY model Green's function. The red dots correspond to a mean field correction of $0.84J_{0}$ ($a=0.84$) to the Larmor frequency (as discussed above), which was determined by optimizing the modulus squared of the overlap between XY and the transverse-field Ising evolutions of $|\Psi_{T}\rangle$ between $tJ_{0}=0$ and $tJ_{0}=1$.

The XY Green's function initially traces the envelope of the fast-oscillating Ising Green's function, but this relationship breaks down at around $tJ_{0}=0.9$. The dots do not reliably track the XY Green's function until $B/J_{0}=15$, but even for $B/J_{0}=20$, the mapping falls off around $tJ_{0}=1$. Further, it is important to note that the gradient of the transverse-field Ising model Green's function at measurement times increases with $B$ because the XY curve does not simply follow its envelope. This means that experimental error will be amplified considerably in the presence of a large transverse field due to timing-jitter errors.

\section{Results}

The plots of the Green's function and of the wavefunction overlap indicate that there is an experimentally optimal field strength that would produce the most accurate simulation of the XY model for a given experimental error in data collection times. This optimal value is important because the slope of the oscillations of the transverse-field Ising model data at the times where data is collected can be huge.  If we assume that an observable oscillates with $\nu \approx \nu_{Larmor} = 2B$, and that $J_{0} \approx 400$~Hz, then the period of oscillation is $1.25 \times \frac{J_{0}}{B}$ milliseconds. A rough calculation of the optimal field strength can be made if we maintain that the experimental error in time measurements must be less than a tenth of the period of the observable. For experimental error of a microsecond, then, $B/J_{0} \leq 125$, for example.

For fields of equal or lower magnitude than the optimal field strength, there are also maximum dephasing times, after which the overlap of the transverse-field Ising model evolved state and the XY evolved state will be too small to say that the two results are equivalent. Note also that the value of the overlap will differ depending on whether a simple Larmor frequency is used or whether a correction factor is included. Table~I summarizes this dephasing time for the modulus squared of the overlap, defining the dephasing time as the time after which the squared overlap is less than $0.7$.

\begin{table*}
\begin{center}
 \begin{tabular}{||c | c | c||} 
 \hline
 Field Strength ($B/J_{0}$) & Larmor Dephasing Time ($1/J_{0}$) & Optimal Dephasing Time ($1/J_{0}$) \\ [0.5ex] 
 \hline\hline
 5 & $0.20$ & $0.28$ \\ 
 \hline
 10 & $0.35$ & $0.79$ \\
 \hline
 15 & $0.50$ & $1.29$ \\
 \hline
 20 & $0.63$ & $1.79$ \\
 \hline
\end{tabular}
\label{tab:table1}
\caption{Dephasing time of the modulus squared of the overlap of the evolved XY and transverse-field Ising states, defined as the time at which $|\langle \Phi_{XY}(t)|\Phi_{Ising}(t)\rangle|^{2} \leq 0.7$ compared to the strength of the transverse field. The data used to determine these values is the same as that presented in Fig.~\ref{fig:overlap}. The Larmor dephasing time corresponds to the black dots in Fig.~\ref{fig:overlap}, which are placed at a frequency of $\omega_{L}=4\pi B$, and the corrected dephasing time corresponds to the red dots, placed at $\omega_{opt}=4\pi \sqrt{B^{2}+(1.67J_{0})^{2}}$.}
\end{center}
\end{table*}

\begin{figure*}
\includegraphics[height=8cm]{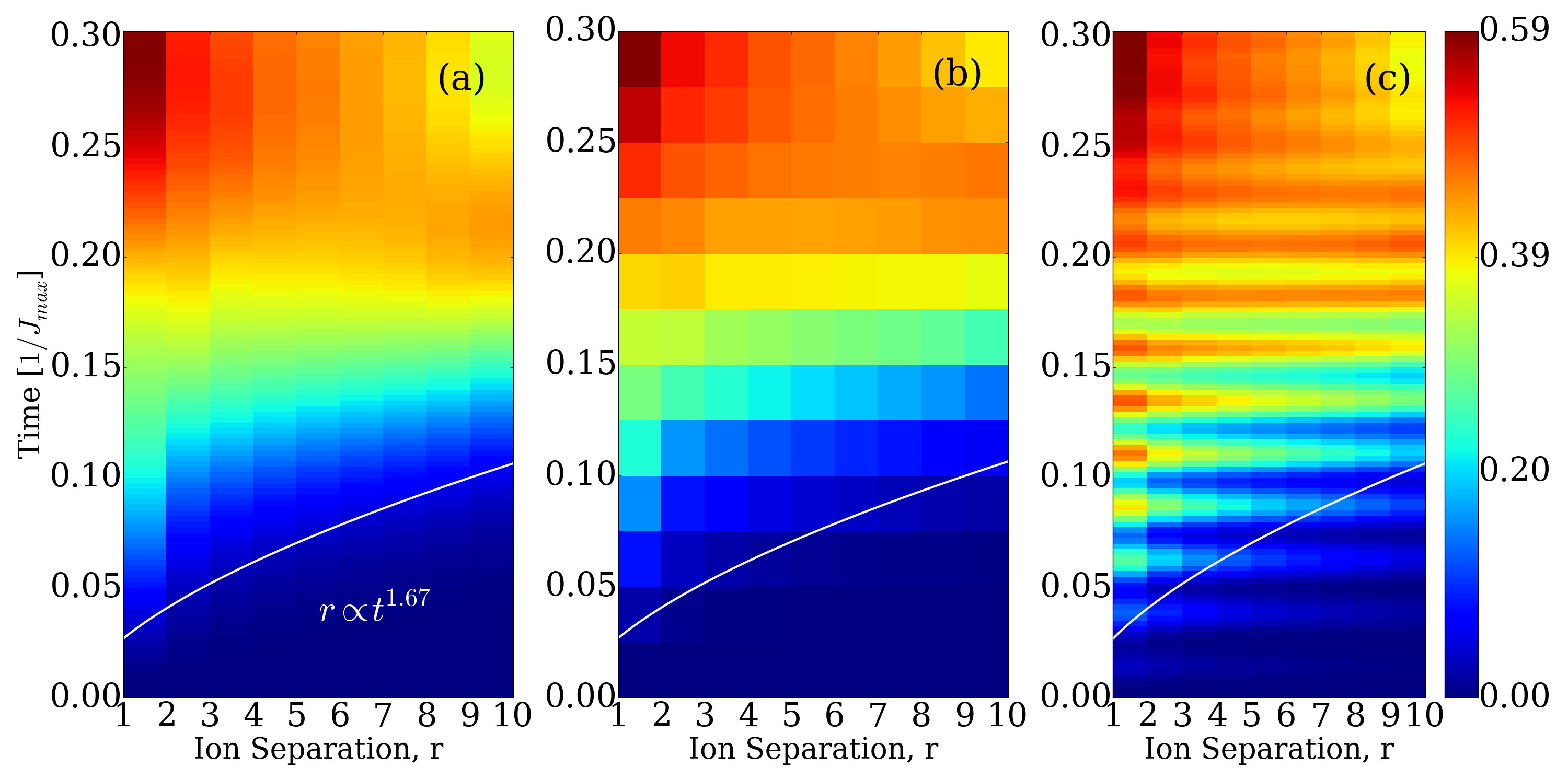}
\caption{(Color online.) Color map for the spatiotemporal evolution of $C_{0,j}(t)$ on an $11$-ion chain with a longitudinal trapping frequency of $560$~kHz ($\alpha\approx 1.19$) and a field strength of $B/J_{0}=10$. Panel (a) plots the evolution of the XY model correlation function between $tJ_{0}=0$ and $tJ_{0}=0.3$. The white curve is a power law fit for the light cone of the correlations, reproduced from Ref.~\onlinecite{monroe_lr}. Panels (b) and (c) plot the evolution of the transverse-field Ising model, but (b) plots only the stroboscopic points with a sampling rate of twice the Larmor frequency, $2\omega_{L}=8\pi B$. Panel (c) samples the system too frequently, so it does not produce the XY model accurately.}
\label{fig:cmaps}
\end{figure*}

This method of simulating an XY model evolution via the rotating-wave approximation has been used  in an evaluation of Lieb-Robinson bounds for propagation speeds in systems with long-range correlations~\cite{monroe_lr}. Their experiment used a Paul trap with $J_{0} \approx  400$~Hz and a transverse field of $B/J_{0} = 10$. They evaluate a static correlation function, $C_{i,j}(t) = \langle \sigma^{y}_{i}(t) \sigma^{y}_{j}(t) \rangle - \langle \sigma^{y}_{i}(t) \rangle \langle \sigma^{y}_{j}(t) \rangle$, between a spin on one end of an $11$-site ion chain ($i=0$) and all other spins in the chain. They also plot the evolution of this function up to $tJ_{0}=0.3$. In Fig.~\ref{fig:cmaps}, we show a numerical evaluation of the same function for a longitudinal trapping frequency of $ 560$~kHz corresponding to $\alpha\approx 1.19$. Their best fit Lieb-Robinson bound is also overlaid on those plots.

Panel (a) gives a numerical evolution of the correlation function for the XY model, while panels (b) and (c) show the transverse-field Ising simulation of the XY model. Panel (b) measures the transverse-field Ising model at the twice the Larmor frequency, which corresponds to the values for which this mapping occurs, while panel (c) measures the Ising model at a frequency eight times greater than the Larmor frequency (four times more frequent than the strobiscpic mapping, as detailed below). Note that the transverse-field Ising model, measured at the appropriate times, provides a good simulation of the XY model over this short timescale. This is not surprising for a field strength of $B/J_{0}=10$, given that the coherence time of the modulus squared of the overlap is $0.35/J_{0}$, or about $9$ milliseconds for $J_{0} \approx 400$~Hz. The white  curve is the power law fit from the experiment. Note that imprecise timing would result in both a qualitatively different color map and an incorrect light cone measurement because the transverse-field Ising oscillations are nonnegligible compared to the features of $C_{i,j}(t)$, even at short timescales. 

Note that the frequency used for the mapping of the correlation function $C_{i,j}(t)$ is $8\pi B$, which is twice that which is used for the overlap and the Green's function. This is because $C_{i,j}(t)$ is dependent upon the operator $\langle\sigma^{y}_{i}(t)\sigma^{y}_{j}(t)\rangle$. When the spins of the $i$th and $j$th ions have made one half rotation in the transverse field, which is oriented in the $\hat{z}$ direction, $\langle\sigma^{y}_{i}(t_{1})\rangle = -\langle\sigma^{y}_{i}(t_{0})\rangle$ and $\langle\sigma^{y}_{j}(t_{1})\rangle = -\langle\sigma^{y}_{j}(t_{0})\rangle$, so $\langle\sigma^{y}_{i}(t_{1})\sigma^{y}_{j}(t_{1})\rangle = \langle\sigma^{y}_{i}(t_{0})\sigma^{y}_{j}(t_{0})\rangle$. Thus, the mapping frequency for $C_{i,j}(t)$ is twice that of a quantity that depends on only one $\sigma^{\alpha}$.

\section{Conclusions}

We examined the mapping between the transverse-field Ising model in a large magnetic field to the XY model in zero field via the rotating wave approximation. We compared the overlap of the wavefunctions for the two models, the time traces of a pure-state Green's function, and a static spin-spin correlation function.  As the field in the Ising model is made larger, the mapping becomes more precise, but the oscillation frequency increases, so the measurement become more susceptible to timing jitter. In addition, objects like Green's functions map to each other only at the precise stroboscopic times, not at the envelope of their values, as occurs in other similar mappings. Finally, if one tries to follow this mapping for too long, it breaks down due to the imprecise mapping period (caused by a finite $B$ field) and due to timing jitter in the measurements. Nevertheless, this mapping can be employed to perform simulations of the more complex XY model for short to intermediate times and is much simpler than directly simulating the full XY model. This approach has already been employed in quantum simulations. We hope our work helps quantify how far one can push this mapping, and more importantly, shows where it fails.

\section{Acknowledgments}
J.K.F. and T.K. acknowledge support from the National Science Foundation under Grant No. PHY-1620555. J.K.F. also acknowledges support from the McDevitt bequest at Georgetown University.

\end{document}